\documentclass[aps,prb,showpacs,twocolumn]{revtex4}
\usepackage{graphicx,amsmath}
\begin{document}
\title{Confinement effects in antiferromagnets}
\author{A.\ D\'{\i}az-Ortiz and J.\ M.\ Sanchez}
\affiliation{Texas Materials Institute, The University of Texas at
Austin, Austin, TX 78712}
\date{\today}
\begin{abstract}
Phase equilibrium in confined Ising antiferromagnets was studied
as a function of the coupling ($v$) and a magnetic field ($h$) at
the surfaces, in the presence of an external field $H$. The ground
state properties were calculated exactly for symmetric boundary
conditions and nearest-neighbor interactions, and a full
zero-temperature phase diagram in the plane $v$-$h$ was obtained
for films with symmetry-preserving surface orientations. The
ground-state analysis was extended to the $H$-$T$ plane using a
cluster-variation free energy. The study of the finite-$T$
properties (as a function of $v$ and $h$) reveals the close
interdependence between the surface and finite-size effects and,
together with the ground-state phase diagram, provides an integral
picture of the confinement in anisotropic antiferromagnets with
surfaces that preserve the symmetry of the order parameter.
\end{abstract}
\maketitle

\section{Introduction}

Confinement effects play an important role in the thermodynamics
of several materials such as polymers, liquid crystals, and
magnets. For example, capillary condensation stands as a well
known example of how phase equilibrium is affected by the
confluence of surface and finite-size effects. In particular, due
to the wall-particle interaction, a fluid between two plates
undergoes a gas-liquid transition at a lower pressure than it does
in the bulk.\cite{fisher1981,nakanishi1983,evans-review-1990,%
binder1992} These effects of confinement are due to the additional
contributions to the thermodynamic potential of the solvation
force (finite-size effect) and the wall-fluid interfacial tension
(surface effect).\cite{evans1987}

A more complicated physical situation arises in the case of thin
films of polymer mixtures on selective
substrates.\cite{binder-review-1999} An $AB$ polymer mixture which
undergoes a phase separation below a bulk critical temperature
$T_c$, develops, when cast into a thin film over a substrate, an
interface between the two phases which runs parallel to the
substrate. This interface appears provided there is a substrate
affinity for one of the components---the confinement is
established between the polymer-air and polymer-substrate
boundaries.

A model fluid confined between two parallel walls that exert opposite surface
fields, has been often considered in order to investigate the underlying
physics in systems with competing boundaries.\cite{brochard1983,parry1990,%
parry1992,swift1991,binder1995-1,binder1995-2,binder1996,ferrenberg1998,%
maciolek1996,rogiers1993,carlon1997,carlon1998} In this case, the interplay
between wetting and phase separation is very important, unlike the case of
capillary condensation in which wetting plays a small role. The competition
between surface effects leads to an interesting and unusual behavior:\ Phase
coexistence is restricted to temperatures below the wetting temperature $T_w$
even in the limit of infinite separation between the plates. The wetting
temperature depends on the surface field and it can be far from the bulk
critical temperature.\cite{wetting} The aforementioned scenario, first
predicted using a mean-field approximation,\cite{brochard1983,parry1990,%
parry1992} has been confirmed subsequently via Monte Carlo simulations\cite{%
binder1995-1,binder1995-2,binder1996,ferrenberg1998} and
transfer-matrix calculations in two dimensions.\cite{maciolek1996}
However, when the effect of gravity is considered phase
coexistence is restored up to the bulk critical
temperature.\cite{rogiers1993,carlon1997,carlon1998}

The confinement studies described above deal with phase separating systems, in
which the phases coexisting along a line of first-order transitions have the
same symmetry, e.g., ferromagnetic thin films. Surface effects in systems with
ordering (antiferromagnetic) interactions have been investigated mostly within
the context of binary alloys undergoing a first-order phase transition,\cite{%
jlml1985,jms1985,mejia1985,sio-mc-review,dosch-review} with particular
emphasis on the surface-induced order and surface-induced disorder
phenomena,\cite{lipowsky1982,lipowsky1987} although some investigations have
been done in the context of multilayer adsorption.\cite{ebner1983,%
kennedy1984,binder1985} More recently, attention has turned to the surface
critical behavior of binary alloys displaying continuous ordering
reactions\cite{dosch-review,schmid1993,drewitz1997,leidl1998,krimmel1997,%
ritschel1996,ritschel1998} and, in particular, to the dependence
of the universality class on surface
orientation.\cite{drewitz1997,leidl1998}

\begin{table*}
\renewcommand{\arraystretch}{1.25}
\caption{Energies for the different ground-states discussed in the
text. Nomeclature is as follows:\ structure
$\uparrow/\uparrow\downarrow/\uparrow$ means that both surfaces are
ferromagnetic (parallel to the applied field) and the remaining $N-2$ inner
layers are antiferromagnetic. Structure $5'$ is a special case in which the 
surface layers are ferromagnetic ($\uparrow$), the subsurface layers are
also ferromagnetic but in the opposite direction ($\downarrow$) and
the rest ($N-4$) are antiferromagnetic ($\uparrow\downarrow$). Bulk and
surface coordination number are denoted by $z$ and $z_s$. See the text for
further explanations.}
\vspace*{3pt}
\begin{tabular*}{\textwidth}{@{\quad}c@{\extracolsep{\fill}}c@{\extracolsep{\fill}}l@{\quad}}
\toprule
Tag&Structure&\multicolumn{1}{c}{Energy}\\
\colrule
1   &$\downarrow/\downarrow/\downarrow$                          &
     ${\mathcal H}_1= z_s+(\frac{1}{2}z+H)(N-2)+2(H+h)$     \\
2   &$\uparrow\downarrow/\downarrow/\uparrow\downarrow$          &
     ${\mathcal H}_2=-z_s+(\frac{1}{2}z+H)(N-2)$              \\
3   &$\uparrow/\downarrow/\uparrow$                              &
     ${\mathcal H}_3=z_s-4z_1+(\frac{1}{2}z+H)(N-2)-2(H+h)$ \\
4   &$\downarrow/\uparrow\downarrow/\downarrow$                  &
     ${\mathcal H}_4= z_s-\frac{1}{2}z(N-2)+2(H+h)$           \\
5   &$\uparrow\downarrow/\uparrow\downarrow/\uparrow\downarrow$  &
     ${\mathcal H}_5=-z_s-\frac{1}{2}z(N-2)$                    \\
$5'$&$\uparrow/\downarrow/\uparrow\downarrow/\downarrow/\uparrow$&
     ${\mathcal H}_{5'}=z_s+z_0-2z_1-\frac{1}{2}z(N-4)-2h$      \\
6   &$\uparrow/\uparrow\downarrow/\uparrow$                      &
     ${\mathcal H}_6= z_s-\frac{1}{2}z(N-2)-2(H+h)$           \\
7   &$\downarrow/\uparrow/\downarrow$                            &
     ${\mathcal H}_7= z_s-4z_1+(\frac{1}{2}z-H)(N-2)+2(H+h)$\\
8   &$\uparrow\downarrow/\uparrow/\uparrow\downarrow$            &
     ${\mathcal H}_8=-z_s+(\frac{1}{2}z-H)(N-2)$              \\
9   &$\uparrow/\uparrow/\uparrow$                                &
     ${\mathcal H}_9=z_s+(\frac{1}{2}z-H)(N-2)-2(H+h)$      \\
\botrule
\end{tabular*}
\label{t1}
\end{table*}

In this paper we investigate the interplay between finite-size and
surface effects in Ising antiferromagnets in the presence of an
external field. In particular, we are interested in systems with
surfaces that preserve the symmetry of the order parameter. In
other words, we shall study thin films which develop
antiferromagnetic (AFM) ordering in each plane parallel to the
surfaces. Our layered system can be described by the following
Hamiltonian:
\begin{multline}
\label{our-ham}
{\mathcal H}=J_b\sum_{ij\not\in\text{surf}}\sigma_i\sigma_j +
J_s\sum_{ij\in\text{surf}}\sigma_i\sigma_j\\
-H\sum_{i\in\text{bulk}}\sigma_i -
(h+H)\sum_{i\in\text{surf}}\sigma_i\,,
\end{multline}
where the spin variable $\sigma_i$ takes the value of $+1$ or $-1$
depending if the spin at site $i$ is pointing up or down, 
respectively. We have assumed that surface sites, in layers 1 and
$N$ for an $N$-layer film, experience a surface field $h$ in
addition to the external magnetic field $H$. On physical grounds, it is
natural to expect that the pair interactions at and near the surfaces
differ from those in the bulk. We approximate the position dependence of
the pair couplings, by allowing the nearest-neighbor intralayer surface
coupling ($J_s$) to differ from the bulk one ($J_b$). Here we
restrict ourselves to case of $J_b>0$ (antiferromagnetic), but we
allow $J_s$ to assume any real value. Also, we specialize
ourselves in the case of localized symmetric surface fields, i.e.,
the field at each surface is the same and acts only at the surface
sites. In the remaining of the paper, the effective pair
interactions, the surface field, and the external magnetic field
($H$) shall be expressed in terms of the bulk AFM coupling
($J_b>0$). The ratio of surface to bulk coupling is then denoted
by~$v$.

\begin{figure*}
\centerline{%
\includegraphics[width=1.8in]{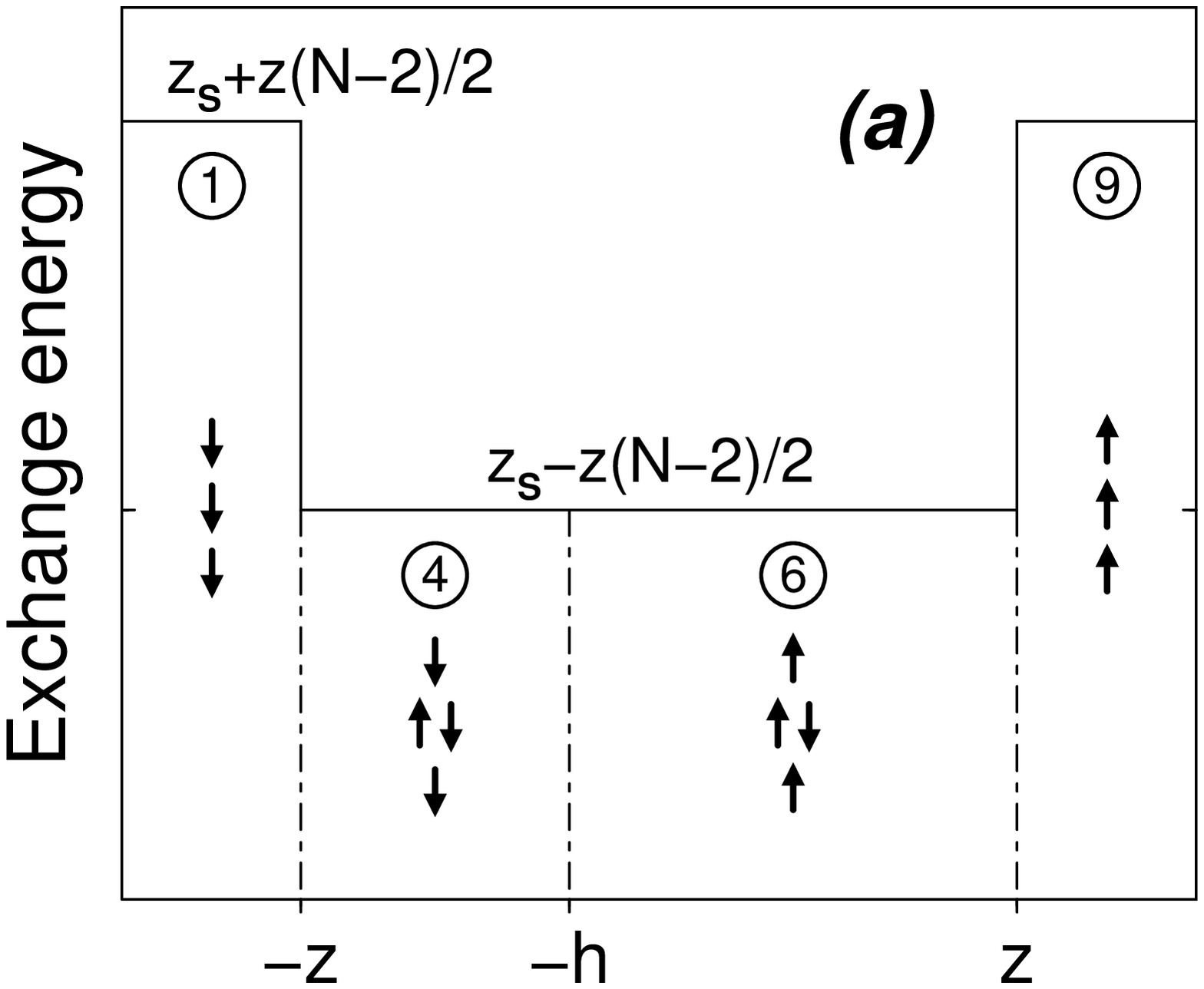}\hfil
\includegraphics[width=1.8in]{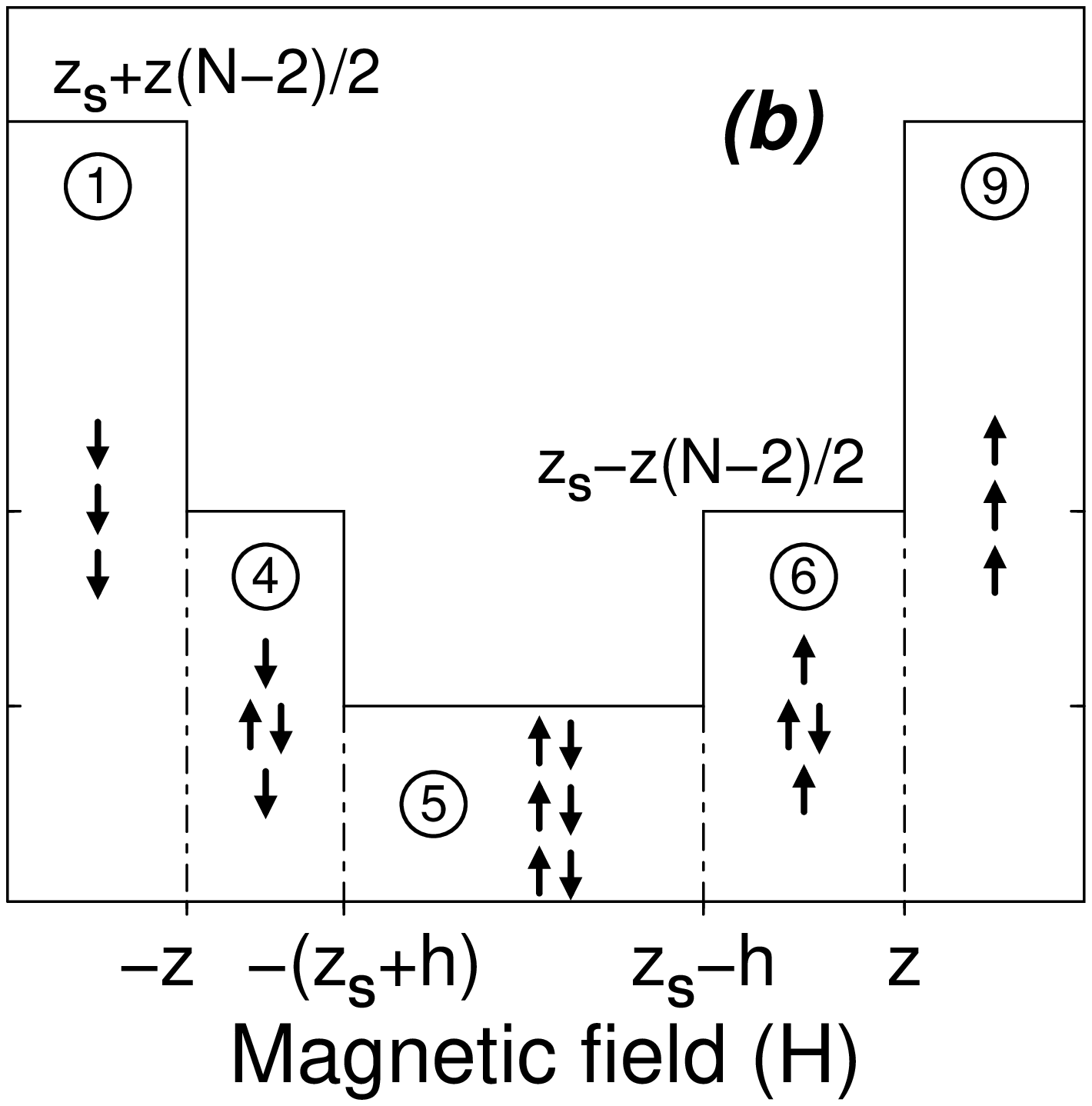}\hfil
\includegraphics[width=1.8in]{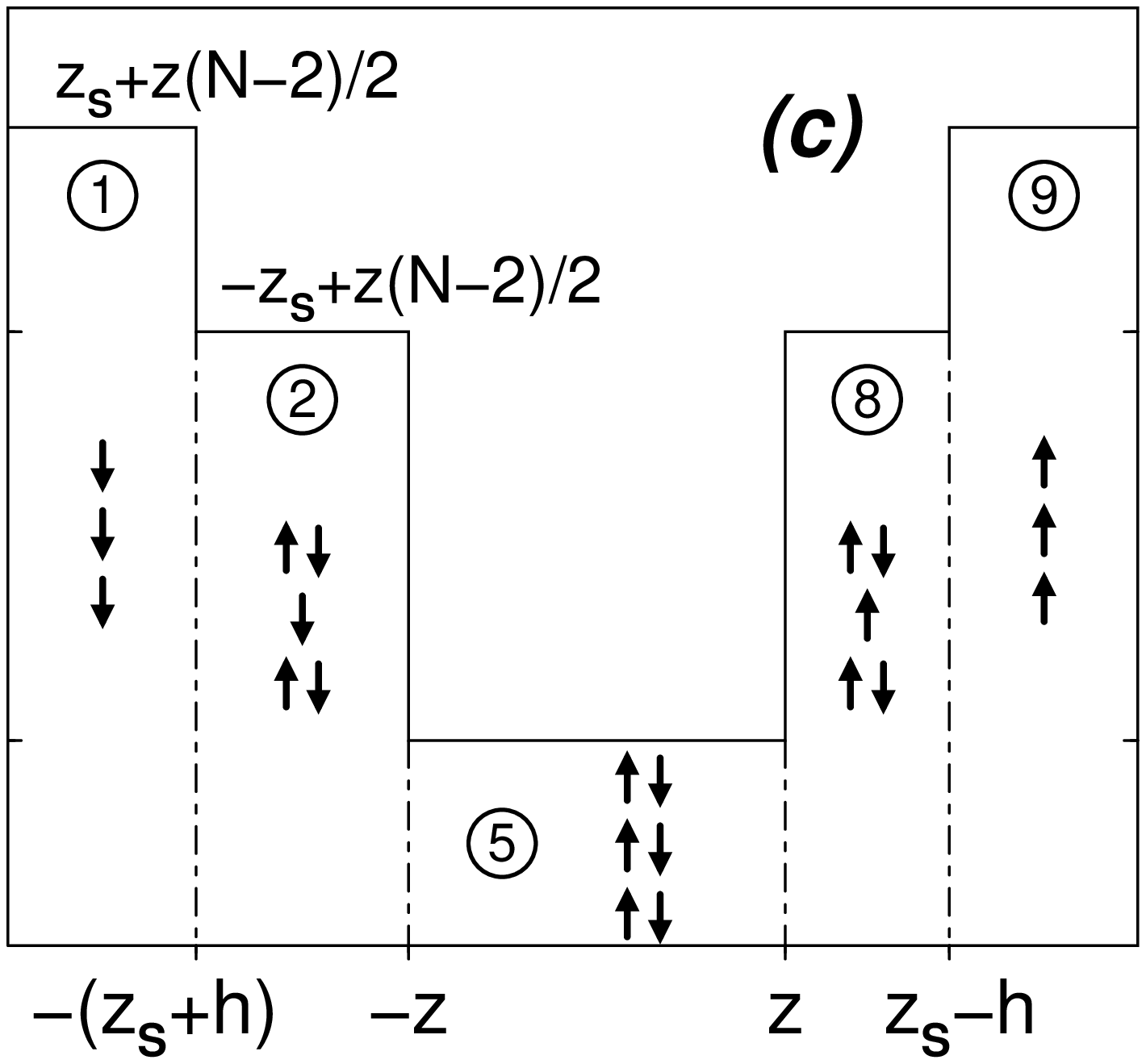}\hfil}\par
\caption{Schematic ground-state sequences VII (a), I (b), and II
(c). The succession (a) to (c) represents the evolution in the
ground-state sequence as we increase the surface pair interactions
for neutral boundary conditions. The domain of stability of each
structure is different from each other, but for $h=0$ the
transition $\mathrm{I\to VII}$ occurs at $v=v_{ps}=-z_1/z_0$ for
which GS 5 in (b) has shrunk to zero width. The range of stability
of GS 5 expands as we increase $v$. At $v=v_m=(z_0+z_1)/z_1$ a
transition between I in (b) and II in (c) occurs. As a reference,
the value of the exchange energy for selected GS is indicated in
the plots. See Fig.\ \ref{f2} for the evolution of sequences I and
II with $h$. Also see the text for further explanations.}
\label{f1}
\end{figure*}

Confinement effects in the order-disorder transitions for the
particular case of $v=1$ and $h>0$ have been reported
previously.\cite{ado-prl-1998} In this paper, we give a full
description of the surface and finite-size effects in terms of the
variables $h$ and $v$. The ground-state properties of the
Hamiltonian (\ref{our-ham}) are derived in Sec.\ \ref{gs}. This
zero-temperature analysis is used to identify the different
sequences of ground states displayed by the film as a function of
the external field $H$. Moreover, it is shown that for
antiferromagnetic systems with symmetry-preserving surface
orientations and nearest-neighbor interactions, a zero-temperature
phase diagram can be drawn as a function of $v$ and $h$, for any
value of the number of layers $N$ and external field $H$. In Sec.\
\ref{ft}, we use a cluster variation free energy\cite{cvm} to
describe the finite-temperature behavior of the system as a
function of surface variables $v$, $h$ and the number of layers
$N$. Particular attention is devoted to the analysis of the
critical curve (in the $H$-$T$ plane) for each one of the
different regions of the zero-temperature phase diagram. We close
with a summary of the important results (Sec.\ \ref{sc}).

\section{Ground-state properties}
\label{gs} 

In the absence of surface and finite-size contributions, that is in the
bulk, the Hamiltonian (\ref{our-ham}) reduces to
\begin{equation}
\label{bulk-ham}
{\mathcal H}_{\text{bulk}}=\sum_{ij}\sigma_i\sigma_j -H\sum_i \sigma_i\,.
\end{equation}
For a two-sublattice antiferromagnet such as body-centered or
simple cubic, the Hamiltonian (\ref{bulk-ham}) has three different
ground states as a function of the external field $H$:\
ferromagnetic ($\downarrow$) for $H<-H_c$; antiferromagnetic
($\uparrow \downarrow$) for $-H_c<H<H_c$ and again ferromagnetic
($\uparrow$) for $H>H_c$. The critical field $H_c$, equal to the
coordination number $z$ [recall that all quantities in Eq.\
(\ref{our-ham}) as well as in Eq.\ (\ref{bulk-ham}) are normalized
to $J_b$], determines the point where the critical curve $T_c(H)$
meets the field axis.

For the AFM thin films studied here [see Hamiltonian
(\ref{our-ham})], the possible ground-state (GS) structures are
listed in Table \ref{t1} along with their corresponding energy. We
considered only the case $h>0$ since the results for $h<0$ can
obtained straightforwardly from the symmetry properties of
Hamiltonian (\ref{our-ham}). The nomenclature in Table \ref{t1} is
as follows:\ structure number 4 corresponds to $\downarrow/
\uparrow \downarrow/ \downarrow$, which means that both surfaces
are ferromagnetic ($\downarrow$) and that the remaining $(N-2)$
inner layers are antiferromagnetically ordered. Structure $5'$, a
special case to be discussed later in the paper, has both surfaces in a
ferromagnetic state ($\uparrow$), the subsurface layers are
ferromagnetic but with magnetization in the opposite direction
($\downarrow$), and the remaining $(N-4)$ layers are
antiferromagnetic.

We arrived at the set of GS in Table \ref{t1} as follows. Since only
nearest-neighbor interactions are included in the Hamiltonian and the
(uniform) surface field acts locally at surface sites, the presence of
long-period superstructures can be ruled out. A possible set of ground states
for Hamiltonian (\ref{our-ham}) was then constructed by combining all possible
surface and bulk ground states. For the sake of definiteness, let us consider
a body-centered cubic film with surfaces in the (110) direction. The bulk
ground states consist of two ferromagnetic structures (with opposite
magnetization) plus an ordered CsCl-type AFM structure. The (110) surfaces
constitute face-centered rectangular lattices, for which the possible ground
states are a checkerboard AFM structure and two ferromagnetic states of
opposite magnetization. The nine ground-state structures obtained by combining
the surface and bulk ground states are listed in Table \ref{t1}. These
structures are ground states of Hamiltonian (\ref{our-ham}) in the limit of
weak coupling between the surface and the subsurface layers. For strong
coupling between the surfaces and the bulk, we found only one additional
ground-state structure---GS $5'$ in Table \ref{t1}.\cite{note1}

The Hamiltonian in Eq.\ (\ref{our-ham}) distinguishes between the pair
interactions in the surface layers from the rest, thus allowing us to define
the surface coordination number $z_s$ as a function of the surface coupling
parameter $v$
\begin{equation}
\label{zs}
z_s(v)=z_0 v + z_1,
\end{equation}
where the intralayer and interlayer coordination are denoted by $z_0$ and
$z_1$, respectively. Recall that all quantities in Eq.\ (\ref{our-ham}) are
given in terms of $J_b$ and, therefore, $z_s$ in (\ref{zs}) actually accounts
for the surface energy. For a bcc(110) film, $z_0=4$ and $z_1=2$, and the bulk
coordination number is $z=z_0+2z_1$.

Even in the absence of an applied surface field $h$, the surfaces are under
the influence of a ``missing neighbors'' field $h_m(v)$ that arises from the
disruption of the translational symmetry perpendicular to the surfaces. This
missing neighbors field produces an inhomogeneous magnetization profile. Thus,
with increasing external field, the surfaces may turn into a ferromagnetic
state before the bulk does. Application of a surface field $h=-h_m(v)$ can
restore the magnetization profile to the homogeneous
condition. Note that the missing neighbors field depends on $v$, since
$h_m$ is a measure of the difference between the environment at the
surfaces and in the bulk [see Eq.\ (\ref{zs})]. The missing neighbors field
can be written as
\begin{equation}
\label{hm}
h_m=z_s-z=z_0v-(z_0+z_1).
\end{equation}
We derive this value for the missing neighbors field later in paper, by
considering the stability of the different GS structures as a function of $v$
and $h$.

A direct comparison between the energies ${\mathcal
H}_i(v,h,H,N)$ for each structure $i$ gives the ground state for every set of
values of the thermodynamic variables (see Table \ref{t1}). However, it is
more useful and less tedious to consider a physical sequence of GS structures
(as a function of the applied field) and examine its domain of stability as we
vary the surface variables $v$ and $h$.

As a starting point, consider the following case:\ Upon the application of an
external field $H$ (in either direction), a film with $v\sim1$ and $h\sim0$
will pass from an AFM state in all layers (small $|H|$) to a state with
ferromagnetic surfaces and an AFM bulk and, finally, for large $|H|$, to a
ferromagnetic state in all planes. This case is represented by the sequence
1-4-5-6-9 of GS structures [a schematic view is presented in Fig.\
\ref{f1}(b). See also Table \ref{t1} for the nomenclature]. The characteristic
value of the external field at the transition between different GS structures
is indicated in Fig.\ \ref{f1}. In general, the transition between GS
structures $A$ and $B$ occurs at ${\mathcal H}_{AB}$, which is determined by
equating the corresponding energies.

Ground-state sequence 1-4-5-6-9 (hereafter referred as I) in Fig.\
\ref{f1}(b), provides some useful insight on confinement versus finite-size
effects. At the beginning of this Section we considered the ground states of
an infinite antiferromagnet, which in the nomenclature of Table \ref{t1},
correspond to GS sequence 1-5-9 in the limit of $N\to\infty$. Thus,
ground-state structures 4 and 6 are due the confinement effects. When either
GS 4 or GS 6 become unstable in favor of GS 5, surface effects are lost and
the film is subject only to the finite-size effects.

An external surface field will produce an asymmetry in the GS sequence
since the Hamiltonian is not invariant under the transformation
$\sigma_i\to-\sigma_i$, $H\to-H$. Applying a surface field $h>0$
reduces the surface ferromagnetism ($\downarrow$) in GS 4 and enhances it
in GS 6 ($\uparrow$). The domain of stability of GS 4 shrinks to zero when
$z_s+h$ becomes $z$. This particular value of $h$ defines the
missing-neighbors field [Eq.\ (\ref{hm})].

Applying an external surface field is not the only way to eliminate
surface effects in AFM thin films. A homogeneous condition can also
be attained in the film by setting neutral boundary conditions
($h=0$) and increasing the pair interactions at the surfaces to a
given value $v_m$. The characteristic value of the surface coupling
that compensates for the missing neighbors effect is given by:
\begin{equation}
\label{vm} v_m=(z_0+z_1)/z_0.
\end{equation}
For this value of the surface coupling GS 4 and GS 6 become unstable
simultaneously [Eq.\ (\ref{vm}) is equivalent to the condition $z_s=z$].

For values of the surface coupling larger than $v_m$, keeping the neutrality at
the boundaries, ordering becomes stronger at the surfaces than in the inner
layers. This situation is represented in Fig.\ \ref{f1}(c). It is worth noting
that the GS sequence 1-2-5-8-9, hereafter referred as II, is stable not only
in the case of $h=0$ but for a range of values of $v>v_m$ and $h$. We will
return to this point later in the paper.

Reducing the surface coupling make surface ordering less stable, until $v$
reaches the characteristic value
\begin{equation}
\label{vps} v_{ps}=-z_1 / z_0,
\end{equation}
for which GS 5 becomes unstable [$v_{ps}$ in (\ref{vps}) corresponds to the
condition $z_s=0$]. The remaining ground-state sequence 1-4-6-9 (hereafter
VII) are depicted in Fig.\ \ref{f1}(a). Sequence VII remains unaltered for
$v<v_{ps}$ regardless the strength of $v$:\ The phase coexistence between
spin-up and spin-down is regulated by the surface field $h$. Large, negative
values of $v$ increase the critical-point temperature. In the alloy
terminology, sequence VII represents the situation of a binary-alloy thin film
with an ordered bulk coexisting with a surface miscibility gap. This will
become apparent in Sec.\ \ref{ft} where we discuss the finite-temperature
properties of Hamiltonian (\ref{our-ham}).

\begin{figure}[t]
\centerline{\includegraphics[width=1.8in]{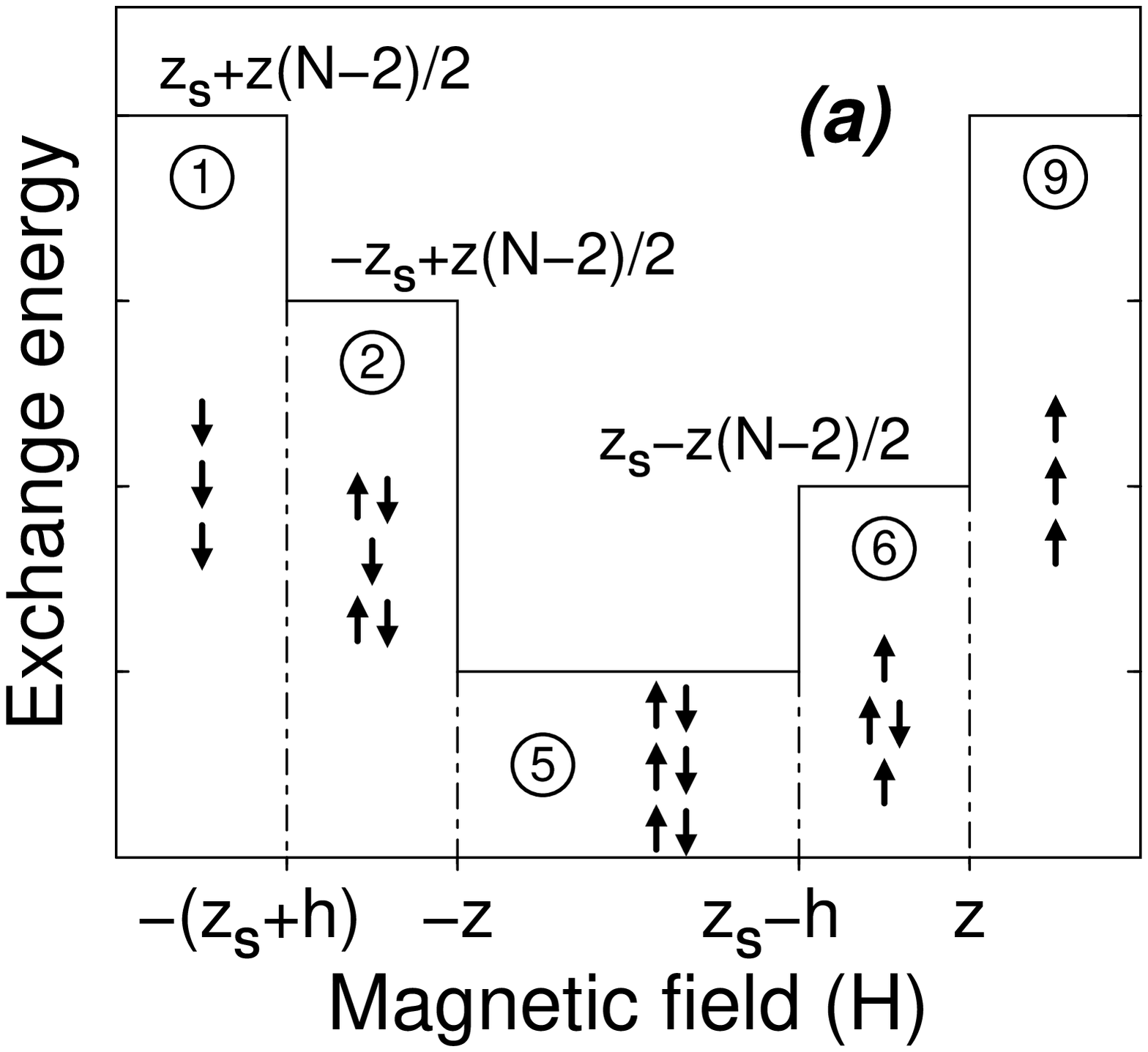}}\par\vskip2\baselineskip
\centerline{\includegraphics[width=1.8in]{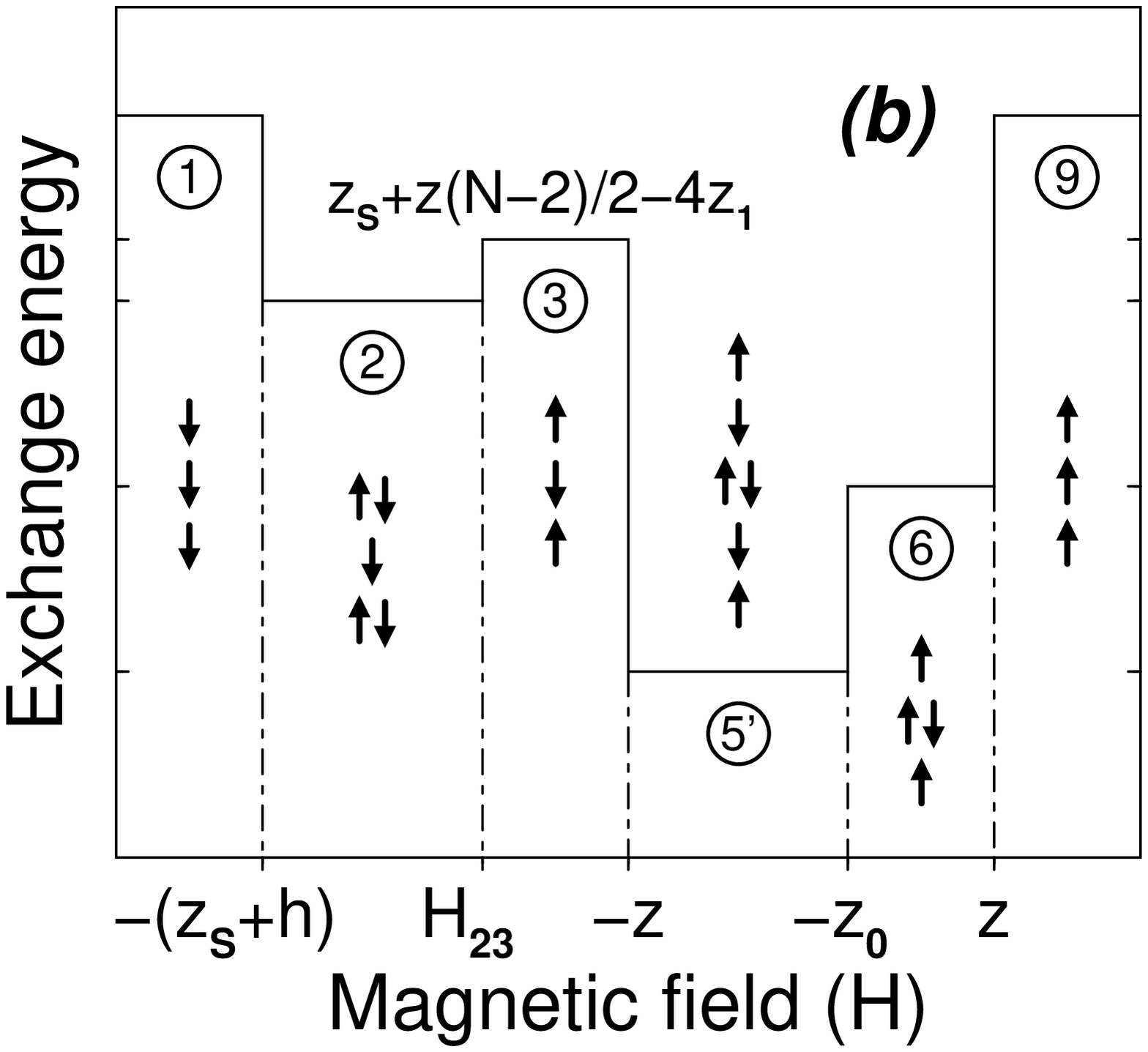}}\par
\caption{Sequence II in Fig.\ \ref{f1}(c) becomes sequence III in
(a) as the surface field is increased. The transition
$\mathrm{II\to III}$ occurs at $h_{\text{II-III}}=z_0v-(z_0+z_1)$
when GS 8 becomes unstable (see Table \ref{t2}). A further
increase of the surface field $h$ establishes 1-2-3-$5'$-6-9 in
(b) as the stable GS sequence (IV). Observe the appearance of GS
$5'$ and the disordered gap between GS 3 and GS $5'$. The
characteristic field between GS 2 and GS 3 is $H_{23}=z_s-2z_1-h$.
See the text.} 
\label{f2}
\end{figure}

Sequences I, II, and VII (Fig.\ \ref{f1}) were obtained by
analyzing the stability of the corresponding GS sequences upon
variations of the surface coupling $v$ for neutral boundary
conditions. As expected, a similar variation of GS sequences will
appear as we increase the surface field. Consider for example
sequence II in Fig.\ \ref{f1}(c):\ Setting higher values for the
surface field eventually overcome the ordering tendencies at the
surfaces. Ground-state structure 8 then becomes unstable and sequence II
turns into the new 1-2-5-6-9 GS sequence (III) depicted in Fig.\
\ref{f2}(a). The asymmetry of sequence III is interesting. For
very negative values of $H$ sequence III looks like sequence II
(in the same range of $H$), with long-range order dictated by the
surfaces. On the other hand, for large positive values of $H$,
sequence III looks like sequence I, for which the bulk is
responsible for the AFM ordering. This similarity is due to the
fact that sequence I evolves into III when the surface field
increases beyond $h=z-z_s(v)$ (missing neighbors field) for
$0<v<v_m$.

The homogeneous antiferromagnetic thin film (GS 5), with constant
energy for given $v$ and $N$, becomes rapidly unstable with
increasing $h$. For sufficiently large $h$, GS 5 is replaced by
another zero-magnetization structure, GS $5'$ in Table \ref{t1},
with energy given by
\begin{equation}
\label{5p}
{\mathcal H}_{5'}=z_s+z_0-2z_1-{\textstyle\frac{1}{2}}z(N-4)-2h.
\end{equation}
For a given value of the number of layers $N$ and the coordination at the
surfaces, ${\mathcal H}_5$ is constant while ${\mathcal H}_{5'}$ depends only
on $h$. When the surface field reaches the value of
\begin{equation}
\label{h14}
h_{\text{III-IV}}=z_0 v+(z_0+z_1),
\end{equation}
structures 5 and $5'$ have the same energy. A unique feature of GS
5 and GS $5'$ is that they remain degenerate over a {\em finite\/}
range of the external field $H$. From Eq.\ (\ref{h14}) and Fig.\
\ref{f2}(a) we can see that the ground state for an AFM film at a
surface field value given by (\ref{h14}) is a {\em mixture\/} of
GS 5 and GS $5'$ for $H\in(-z,-z_0)$.  On average, a scan in $H$
will show a layer magnetization of $+\frac{1}{2}$ at the surfaces
together with subsurface magnetization of $-\frac{1}{2}$ and AFM
bulk (zero magnetization). Thermal excitations destroy this
degeneracy between GS 5 and GS $5'$ in most of the interval
$(-z,-z_0)$ in favor of GS 5, except near the ends, i.e.\
$H\sim-z$ and $H\sim-z_0$, where GS $5'$ is pinned by the onset of
stability of GS 3 and the presence of GS 6. Traces of the
ground-state degeneracy between GS 5 and GS $5'$ are observable at
low temperatures. For the other structures listed in Table
\ref{t1}, a transition similar to $5\to5'$ does not occur, mainly
due to the symmetry in the boundary conditions.

Ground-state sequence III evolves into 1-2-3-$5'$-6-9 sequence (IV
hereafter) at $h=h_{\text{III-IV}}$. The situation is shown
schematically in Fig.\ \ref{f2}(b) for $h>h_{\text{III-IV}}$.
Observe that the appearance of GS 3 has established a disorder gap
between GS 2 (AFM surfaces) and GS $5'$ (AFM bulk). This behavior
is unique in the sense that in all previous cases the ordered
domain was a compact interval in $H$. This characteristic brings
some interesting features into the $H$-$T$ phase diagram, such as
the splitting of the film's critical $T_c(H)$ curve into two
distinct critical curves.\cite{ado-prl-1998}

\begin{figure}[t]
\centerline{\includegraphics[width=1.8in]{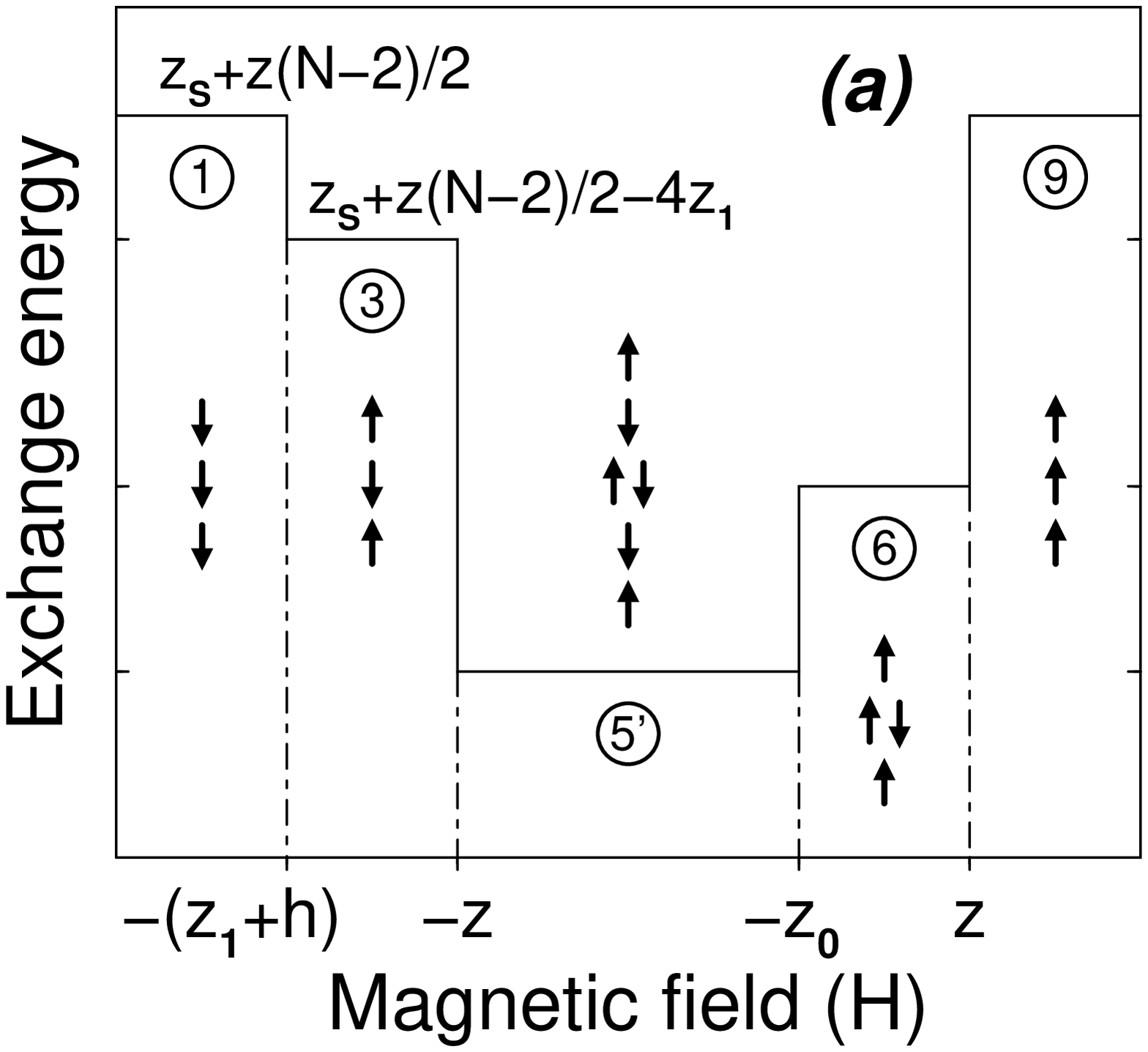}}\par\vskip2\baselineskip
\centerline{\includegraphics[width=1.8in]{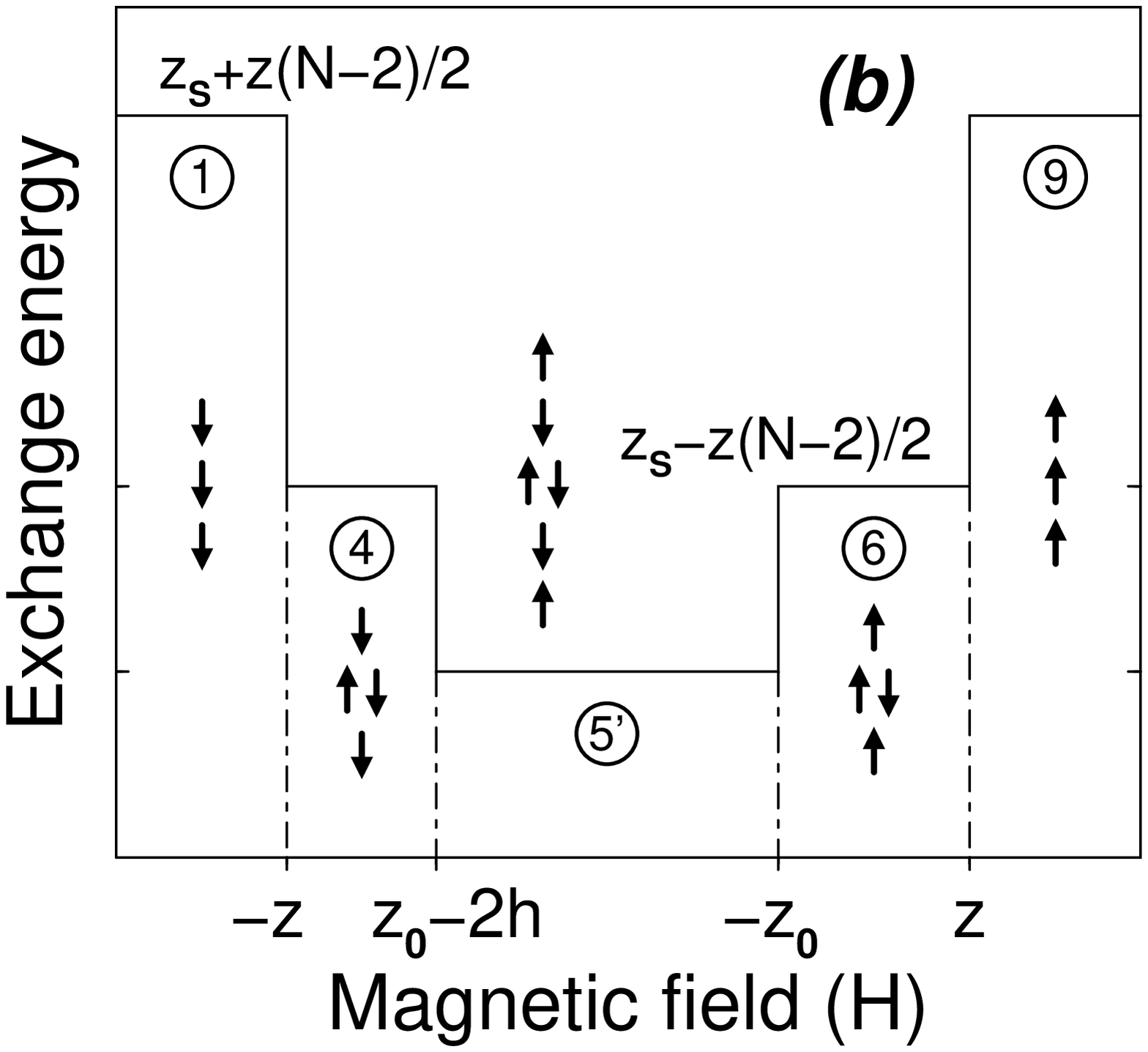}}\par
\caption{Ground-state sequences for intense surface fields and
negative $v$. Sequence V in (a) is obtained from IV in Fig.\
\ref{f2}(b), when the surface AFM-phase in GS 2 becomes unstable
upon the reduction the pair interaction at surfaces. A coexistence
between spin-up (GS 3) and spin-down (GS 1) magnetizations is
established at the surfaces [{\em cf}.\ Fig.\ \ref{f1}(a)]. For
sufficient negative values of $v$ and large $h$, a line of
first-order transitions, ending at a critical point, occurs
outside the antiferromagnetic critical curve. Sequence I in Fig.\
\ref{f1}(c) turns into sequence VI in (b) in the same way as III
becomes IV (Fig.\ \ref{f2}), that is, replacing GS 5 by GS $5'$. A
difference arises, however, since in this case GS 4 is adjacent to
GS $5'$, and a coexistence line will appear {\em inside\/} the AFM
region. See the text for details.} \label{f3}
\end{figure}

Increasing the surface field does not change sequence IV into another GS
sequence. However, for a large value of $h$, antiferromagnetic order at the
surfaces becomes unstable upon reduction of the surface coupling, and IV
changes into sequence V composed of GS sequence 1-3-$5'$-6-9  for $v<0$ [Fig.\
\ref{f3}(a)]. Observe in Fig.\ \ref{f3}(a) that GS 1 is now adjacent to GS
3. The difference between GS 1 and GS 3 resides at the surfaces, which have
opposite magnetization. This situation is reminiscent to the one found in
sequence VII [Fig.\ \ref{f1}(a)] where the surface field regulates the surface
phase coexistence between up- and down-magnetizations. In sequence V, however,
the line of first-order transitions is located outside of the well defined AFM
region composed by GS $5'$ and GS 6.

Ground-state sequence V is stable for large $h$ and negative $v$.
Previously, we found that VII is the stable GS sequence for
$v<v_{ps}$ and low $h$. A transition between V and VII certainly
occurs, although it is mediated by the GS sequence 1-4-$5'$-6-9
(VI) [see Fig.\ \ref{f3}(b)]. Finally, sequences I and VI are
separated by the GS 5 to GS $5'$ transition, which in this case
occurs at lower values of $h$ since $v<0$.

\begin{figure*}
\centerline{\includegraphics[width=2.0in]{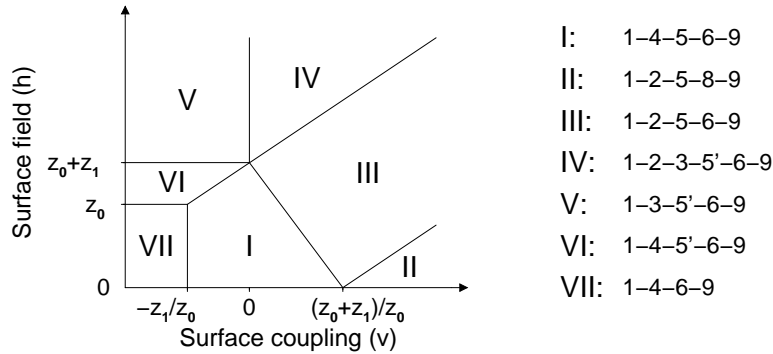}}\par
\caption{Ground-state phase diagram $v$-$h$ for antiferromagnetic
thin films. Table \ref{t2} contains the characteristic values of
the surface field that describe the boundary between the different
regions. Regions I--III define compact antiferromagnetic domains,
while for region IV a disordered (ferromagnetic) gap intervenes
between the AFM order at the surfaces and the ordered bulk. For
large negative values of the surface coupling, a line first-order
transitions ending in a critical point, appears in regions V--VII.
The intra- and interlayer coordination numbers are denoted by
$z_0$ and $z_1$, respectively. See the text for further
explanations.} \label{gspd}
\end{figure*}

We have discussed the several GS sequences that appear in confined
antiferromagnets along particular paths, namely, we fixed $h=0$ and
varied the surface coupling (Fig.\ \ref{f1}) or alternatively, we
fixed $v>v_m$ and increased $h$ (Fig.\ \ref{f2}). In general,
however, the transition from one GS sequence to another does not occur at
constant $v$ or $h$. The relationship between the surface variables $(v,h)$
at the different transitions between GS sequences (see Table \ref{t2}),
defines the domain of stability of each sequence, from I to VII,
in the plane $v$-$h$. The corresponding ground-state phase diagram
is shown in Fig.\ \ref{gspd}. The phase diagram is symmetric with
respect to $h=0$, with the negative-$h$ region obtained by
replacing spin-up with spin-down and $H$ with $-H$ in Fig.\
\ref{gspd}. The zero-temperature phase diagram provides a good
reference frame to interpret some of the features reported in
previous work on binary-alloy thin films with ordering
interactions.\cite{ado-prl-1998,elisa1995,ado-cms-1997,ado-ssc-1998}
The ground-state phase diagram is also a valuable guide for the
investigation of the finite-temperature properties of Hamiltonian
(\ref{our-ham}) to be carried out in the next Section.

\section{Finite-temperature properties}
\label{ft}

The finite-temperature properties of Hamiltonian (\ref{our-ham}) were
calculated using the cluster-variation method (CVM) in the pair approximation
(PA).\cite{cvm} For the two-sublattice antiferromagnets considered in this
paper, the physical aspects of phase equilibrium under confinement are well
captured by the PA-CVM.\cite{ado-prl-1998,ado-ssc-1998} For bcc(110) films
with neutral boundary conditions, a comparison between the PA and the
tetrahedron approximation (TA) has shown that only the quantitative aspects
are improved with the latter.\cite{ado-ssc-1998} For a general exposition of
the cluster-variation method, we refer the interested reader to the excellent
reviews available in the literature.\cite{ducastelle,cvm-ddf,cvm-finel,%
cvm-ppm1,cvm-ppm2}

The order-disorder transitions are described in the usual manner by
subdividing the bcc or sc lattice into two interpenetrating sublattices
$\alpha$ and $\beta$. The long-range order parameter in the $k$-layer defined
as
\begin{equation}
\label{lro}
\eta_k={\textstyle\frac{1}{2}}(m_\alpha^k - m_\beta^k) ,
\end{equation}
where $m_{\alpha(\beta)}^k$ is the $\alpha (\beta)$-sublattice magnetization
in the $k$ layer.

\begin{table}[h]
\caption{Characteristic values of $h$ and $v$ defined by the 
transitions between different ground-state sequences. These characteristic
values define the domain of stability of structures I to VII. See phase
diagram in Fig.\ \ref{gspd} and the text for additional details.}
\vspace*{3pt}
\begin{tabular*}{\columnwidth}{@{\hspace*{50pt}}l@{\extracolsep{\fill}}}
\toprule
\multicolumn{1}{c}{\hspace*{50pt}Surface field/coupling}\\
\colrule
$h_{\text{I-III}}=-h_{\text{II-III}}=-z_0 v+(z_0+z_1)$\\
$h_{\text{I-VI}}=h_{\text{III-IV}}=z_0v+(z_0+z_1)$\\
$h_{\text{V-VI}}=z_0+z_1$\\
$h_{\text{VI-VII}}=z_0$\\
$ v_{\text{I-II}}=(z_0+z_1)/z_0$\\
$v_{\text{I-VII}}=-z_1/z_0     $\\
\botrule
\end{tabular*}
\label{t2}
\end{table}

\begin{figure}
\centerline{\includegraphics[width=2.2in]{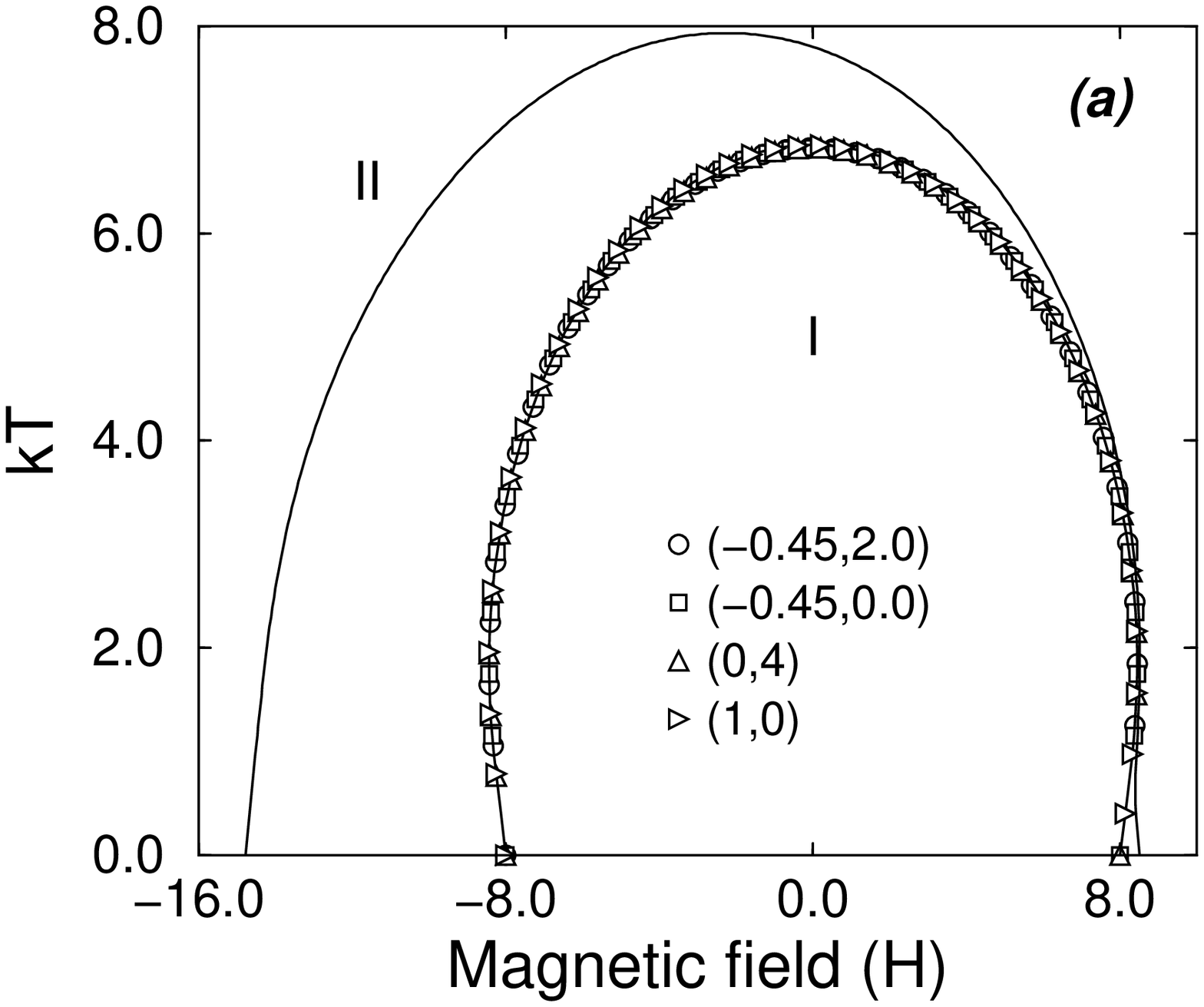}}\par
\centerline{\hspace*{3mm}\includegraphics[width=2.3in]{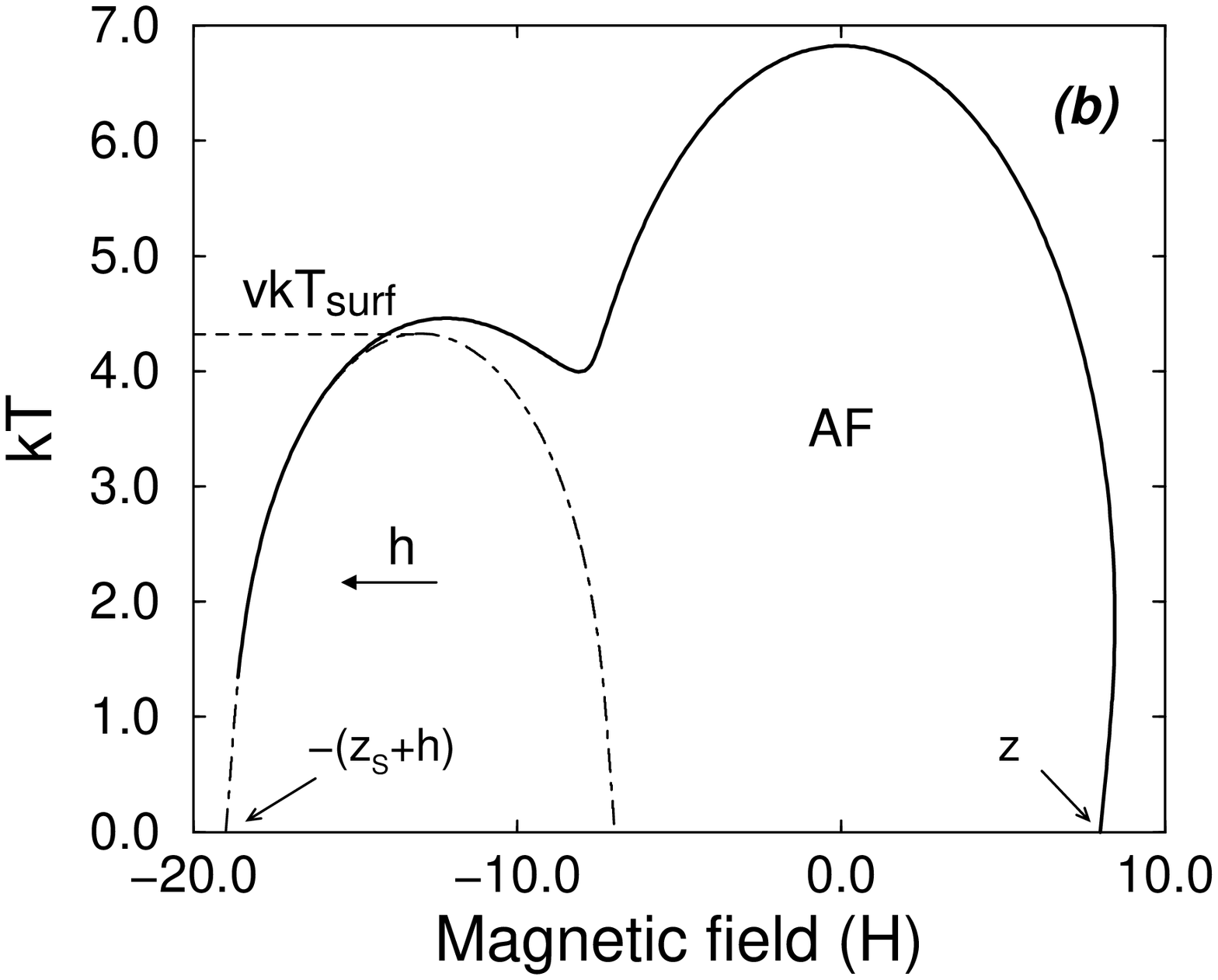}}
\vskip-\baselineskip 
\caption{(a) Typical critical curves for regions I (symbols) and II (solid
line, $v=2.41$ and $h=3.14$). The various symbols correspond to different
values of the surface variables $(v,h$) as shown in the inset. Note that in
region I the shape of the critical curve is virtually independent of $v$
and $h$. (b) Phase diagram for region III (solid line), showing the
development of a `shoulder' as a signature of the incipient surface
critical curve. The values for the surface variables are $v=1.5$ and
$h=11$. The antiferromagnetic domain is a compact region. The phase diagram 
for a square lattice is shown as reference (dot-dashed line) and to
illustrate the process of separation between the bulk and surface critical
curves ({\em cf}.\ Fig.\ \ref{ft2}). Both in (a) and (b) antiferromagnetic
bcc(110) films with $N=14$ were considered and solved in the pair
approximation of CVM.}
\label{ft1}
\end{figure}

With reference to the GS phase diagram of Fig.\ \ref{gspd}, regions I--III
display long-range order, either at the surfaces (GS 2 and 8) or in the bulk
(GS 4--6). With the exception of sequence I, the critical
curves\cite{note2} obtained in regions II and III show a distortion at high
temperatures. Our results for the critical curve in these regions,
summarized in Fig.\ \ref{ft1}, can be explained using the ground-state
analysis discussed in Sec.\ \ref{gs}.

Phase diagrams in region I are virtually independent of the
parameters $v$ and $h$, as can be seen in Fig.\ \ref{ft1}(a). This
behavior can be attributed to the fact that the AFM ordering in
region I is primarily due to the inner layers [see Fig.\
\ref{f1}(b) and Fig.\ \ref{gspd}]. Thermal excitations can promote
spin flip at the surfaces, resulting in a lower degree of ordering
at surfaces relative to the bulk.  In contrast, region II is
characterized by a strong AFM ordering at the surfaces coupled
with the AFM bulk [see Fig.\ \ref{f1}(c)], thus preventing the
formation of a (separate) surface critical curve. Instead, the
$H$-$T$ phase diagram shows an increase in the transition
temperature and a broadening in the external field region for
which the stable phase is antiferromagnetic. A relative small
asymmetry in the critical curve is observed, due to the fact that
the surface field favors the stability of GS 2 over GS 8 [Fig.\
\ref{ft1}(a)]. Thus, the distortion in the phase diagrams
associated with region II stems from the relative stability of two
ground-state configurations with the same symmetry, i.e., GS 2 and
GS 8.

A higher asymmetry in the phase diagrams is expected in region
III, since the critical-curve shape is dictated by the surface
ordering for $H\sim -(z_s+h)$, and by an AFM bulk (with low
surface ordering) for $H\sim z$. The difference in symmetry of the
AFM structures at each AFM:FM boundaries [see the GS sequence in
Fig.\ \ref{f2}(a)], allows the surfaces to drive the phase
transition for fields close to the (negative) critical field
value. One can see that the surfaces are developing their own
critical curve, which unfolds as a `shoulder' in the phase diagram
for negative applied field [see Fig.\ \ref{ft1}(b)].
Characteristics such as the maximum temperature of the shoulder or
its extension in $H$, are controlled by the surface
variables $v$ and $h$. The critical field between GS 1 and GS 2
[$H_{12}=-(z_v+h)$] makes apparent that the extension of the
shoulder depends on the surface field. The maximum temperature in
the shoulder is about $vT_{\text{surf}}$, where $T_{\text{surf}}$
is the N\'eel temperature of the corresponding surface
antiferromagnet. Here, as in the rest of the paper, the relevant
thermodynamic variables are expressed in units of the (positive)
AFM coupling. Thus, in the PA a square lattice has a maximum
critical temperature $kT_{\text{surf}}=4/\ln4\approx2.88$.

\begin{figure}
\centerline{\includegraphics[width=2.3in]{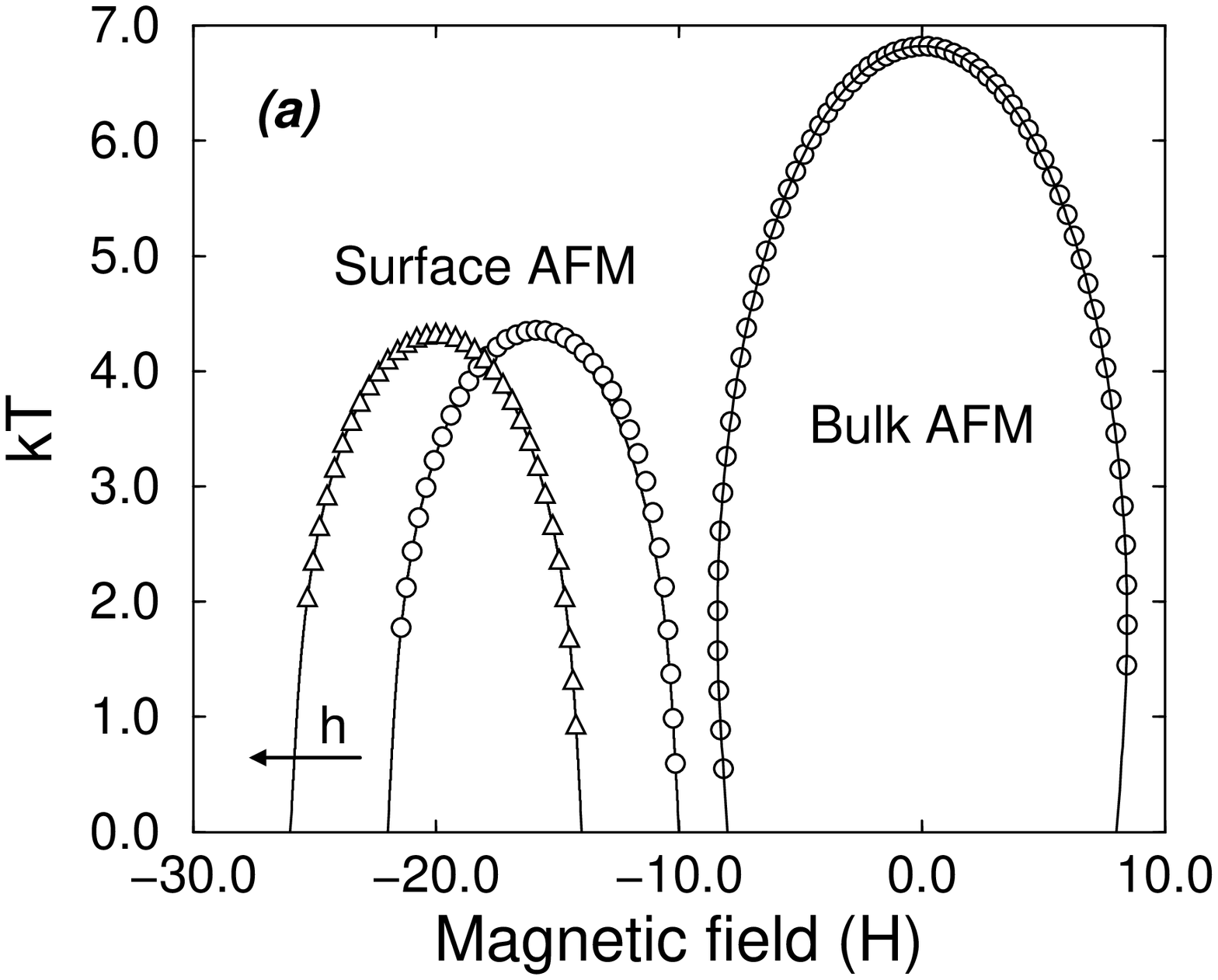}}\par
\centerline{\includegraphics[width=2.3in]{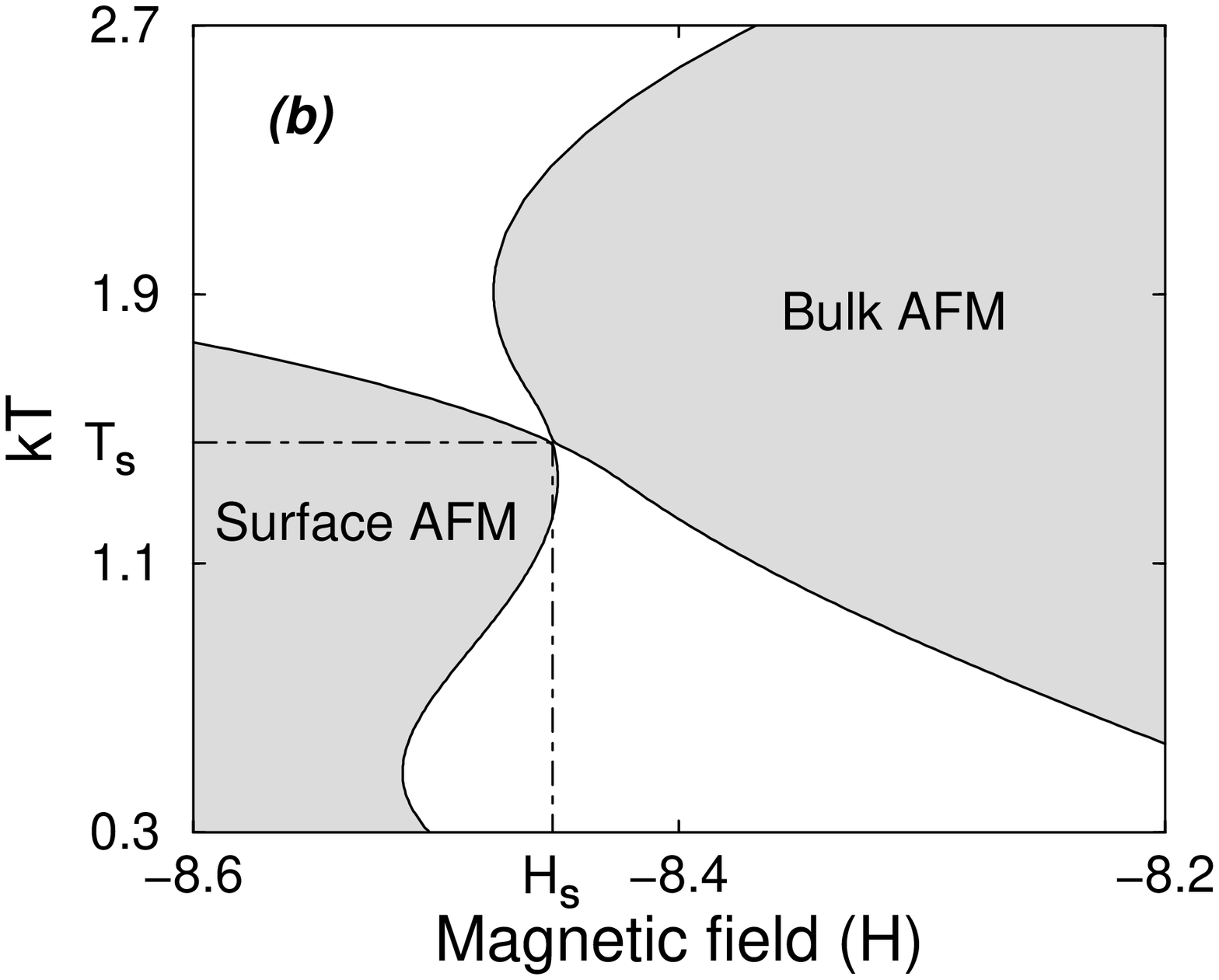}}\par 
\vskip-\baselineskip
\caption{For an intense field at the boundaries ($h>h_s$), the surfaces
decouple from the inner layers (bulk) and develop their own critical
curve. In (a) the critical curves of 14-layer antiferromagnetic thin films
are shown for $v=1.5$, $h=14$ (circles) and $v=1.5$, $h=18$
(triangles). The bulk critical curve showed no difference from $h=14$ to
$h=18$, hence only the former case is depicted. The solid lines represent,
in the case of the surface critical curves, the phase diagram of a square
AFM, appropriately shifted. The solid line in the bulk phase is the one
associated to $N=12$ in region I. The splitting between the surfaces and
bulk critical curves occur at $T_s$ (temperature of splitting) and $H_s$
(field of splitting), when the surface field reaches the value of
$h_s$. Part (b) shows a detail of the phase diagram of 100-layer AFM film
at the very point of splitting.}
\label{ft2}
\end{figure}

As pointed out previously, region IV is characterized by the
formation of a disordered gap between two different ground states
[see Fig.\ \ref{f2}(b)]. At finite temperatures and deep inside
region IV, the surfaces develop their own critical curve well
separated from the bulk antiferromagnetic region [see Fig.\
\ref{ft2}(a) showing the critical curves for a 14-layer film with
$v=1.5$, $h=14$ (circles) and $h=18$ (triangles)]. Since the
surfaces are weakly coupled with the bulk, the surface critical
curve scales with $v$, i.e., the zero-temperature width of the AFM
ordering is $z_0 v$ and the maximum critical temperature is
$vkT_{\text{surf}}$.

Between the situation of unconnected ordered domains and the phase
diagrams observed in region III, there is the case in which the
zero-temperature disordered gap transforms, via thermal excitations, into a
disordered region in the $H$-$T$ plane right inside the compact AFM domain.
An increment in the surface field translates into an increment in
the height of the disordered region. At $h=h_s$ the AFM region
splits into the surface and the bulk critical curves [see Fig.\
\ref{ft2}(b)]. At finite temperatures, the splitting value of the
surface field $h_s$ plays the role of $h_{\text{III-IV}}$:\ for
$h<h_s$ the ordered region is compact whereas for $h>h_s$ there
are two unconnected critical curves.

Expressing the free energy $F$ in terms of the long-range order
parameters (\ref{lro}), the conditions determining the locus of
the splitting point are given by:
\begin{subequations}
\label{split}
\begin{equation}
\lambda=\det\biggl(\frac{\partial^2 F} {\partial \eta_k\,\partial
\eta_{k'}}\biggr)=0,\label{split:hess}
\end{equation}
\begin{equation}
\frac{\partial\lambda}{\partial T}=0,\qquad
\frac{\partial\lambda}{\partial h}=0.\label{split:hess-ht}
\end{equation}
\end{subequations}
Equation(\ref{split:hess}) defines the critical temperature, at
fixed external conditions ($T$, $H$, $h$ and $N$), when the second
derivatives of the free energy are evaluated in the disordered
state.\cite{jms1978} Since $\lambda<0$ in the ordered state, at
the splitting point ($T=T_s$) $\lambda$ is a concave function of
the external field vanishing at the splitting value of the
magnetic field $H_s$. In a similar fashion, one can see that
$\lambda$ is a convex function of temperature, becoming zero at
$T=T_s$ [see Fig.\ \ref{ft2}(b)]. Thus, the splitting point is
defined as a saddle point of $\lambda$ in the $T$ and $H$
variables. Conditions (\ref{split:hess-ht}) account for this.

\begin{figure}
\centerline{\includegraphics[width=2.3in]{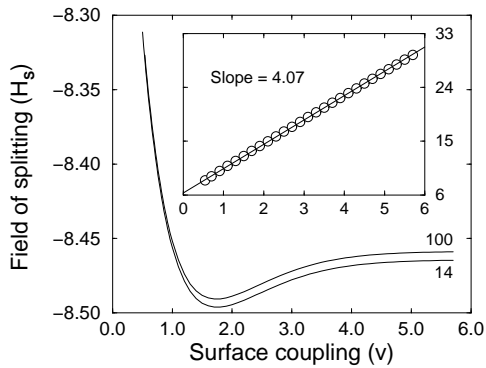}}
\caption{Splitting field $H_s$ as a function of the surface
coupling for AFM bcc(110) films with $N=14$ and $N=100$. The
minimum of $H_s$ is due to the reentrance at low temperatures of
the bulk critical curve. Inset:\ Splitting value of surface field
$h_s$ as a function of the surface coupling $v$, for the case of
$N=100$ (circles). A least-squares fit (solid line) gives
$h_s=4.07v+6.36$. Compare this with $h_{\text{III-VI}}=4v+6$
obtained in Sec.\ \ref{gs} for the boundary between regions III
and IV. See the text for further details.} \label{ft3}
\end{figure}

Using conditions (\ref{split}) we determined the splitting value
of the external field $H_s$ as a function of the surface coupling
for thin ($N=14$) and thick ($N=100$) films. The results are shown
in Fig.\ \ref{ft3} for the case of bcc(110) films. The particular
shape of $H_s(v)$ can be understood as follows:\ Since the height
of the critical curve associated with the surfaces scales with $v$
[see for example Figs.\ \ref{ft1}(b) and \ref{ft2}(a)] and because
of the reentrance of the bulk critical curve, for small $v$ the
point of contact (splitting) between the two critical curves is
shifted to higher values of $H$. As we increase the surface
coupling, the splitting point moves (clockwise) along the bulk
critical curve, reaching a minimum in $H$ and increasing again
towards the saturation value.

We found that within the PA the minimum in $H_s(v)$ is not very
sensitive to the total number of layers. For bcc(110),
$H_s^{\text{min}}$ occurs at $v\sim1.74$ while for sc(100) the
$H_s$ is minimum at $v\sim1.2$. Again, this can be explained by
considering the different N\'eel temperature values for sc and bcc
lattices. The ratio between the latter and the former is $\sim$1.4
(PA), which is comparable to the ratio of the corresponding
$H_s^{\text{min}}$ ($\sim$1.45). The behavior of the other
quantities of interest can be inferred from Fig.\ \ref{ft3}. The
most interesting part, however, is contained in the inset of Fig.\
\ref{ft3}, which shows $h_s$ as a function of the surface coupling
$v$. A least-square fit gives $h_s^{\text{bcc}}=4.07v+6.36$ which
is almost parallel (and very close) to $h_{\text{III-IV}}$ in Eq.\
(\ref{h14}). For sc(100) films similar results were obtained and a
linear fit for $h_s$ gives $h_s^{\text{sc}}=4.04v+5.20$. Thus, the
process of splitting occurs within a narrow interval of $h$.

\begin{figure}[t]
\centerline{\includegraphics[width=2.3in]{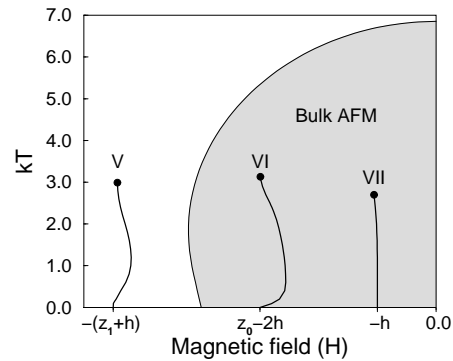}}
\caption{A line of first-order transitions (thick line) appears in
each of the regions V--VI for negative values of the surface
coupling. In all cases, the coexistence is between up- and
down-ferromagnetism at the surfaces. In all cases the bulk
critical curve is not affected by the presence of the first-order
transitions. Since the antiferromagnetic domain (shaded region) is
symmetric around $H=0$ only the left half is shown. The
calculations were done in the PA-CVM for the following values of
the surface variables:\ $v=-1$ and $h=2$ (VII); $h=5$ (VI); $h=9$
(V).} \label{ft4}
\end{figure}

Due to the equivalence between the N\'eel point and the critical
point of a ferromagnet in zero field, the finite-temperature
behavior of AFM thin films, as a function of the surface coupling
$v$ and $H=h=0$, is equivalent to the multicritical phenomena
occurring at the surface of semiinfinite
ferromagnets.\cite{binder-ptcp}  In our case, negative surface
pair interactions give rise to a line of first-order transitions
in regions V--VII (see Fig.\ \ref{ft4}). In all cases the
coexistence line separates surface ferromagnetic phases
with opposite magnetization, that have the same
symmetry. The bulk, however, may have different symmetry at each
side of the coexistence line, thus modifying the shape of the
first-order line at finite temperatures. This can be observed in
Fig.\ \ref{ft4}, where the surface coexistence curve is drawn
$v=-1$ and $h=2$ (VII), $h=5$ (VI), $h=9$ (V). In each case, the
coexistence curve ends in a critical point which is close, as
expected, to the Curie point associated with the (2D) surface
lattice, i.e.\ $\sim |v|T_c$. In all the three regions V--VII, the
AFM bulk remains undisturbed by the presence of the surface
coexistence line. At $v=v_N$ and $H=h=0$ the critical end point
reaches the second-order critical curve at the N\'eel temperature
$T=T_N$. The multicritical behavior is the (trivial) superposition
of two independent critical behaviors which do not interfere with
each other.\cite{binder1985}

\section{Summary and conclusions}
\label{sc}

In this paper we performed an analysis of the confinement effects on
antiferromagnets with symmetry-preserving surface orientations. The
ground-state properties of the model, an Ising Hamiltonian with nn-pair
interactions in the presence of external bulk and surface fields, shows an
interesting structure. A zero-temperature phase diagram in the surface
variables $v$ (surface coupling) and $h$ (surface field) was obtained for
two-sublattice antiferromagnets. In this case there are seven different
regions in the ground-state phase diagram. Each region is characterized by a
particular sequence of ground states as a function of the external field. An
analysis of the ground-state phase diagram explains (and sometimes even
anticipates) some of the features found in the $H$-$T$ critical
curves. Together with an examination of the finite-temperature behavior in
each of the aforementioned regions, our analysis showed that the interplay
between the surface variables $v$ and $h$ defines the thermodynamics of
confinement in ordering systems. For example, the splitting of the critical
curve into surface and bulk contributions results from the simultaneous
application of seemingly competing contributions $v>v_m$ (ordering) and
$h>h_{\text{II-IV}}$. At the other extreme, the development of a surface
coexistence line for $v<0$ and $h>0$ represents a particular case of magnetic
surface reconstruction.

\acknowledgments
A.D.-O gratefully acknowledges the financial support from CONACyT through
the Post Doctoral Fellowships Program and under grant G-25851-E.



\begin{thebibliography}{99}
\bibitem{note1} We confirmed these results by evaluating the free energy (in
the pair approximation) at low temperatures, as function of the surface
variables $h$ and $v$ for different values of the  external field $H$. We did
not find any other ground-state structure additional to those listed in Table~\ref{t1}.
\bibitem{note2} At finite temperatures, the antiferromagnetic region
[nonzero value of the order parameter in (\ref{lro})] is separated
from the disordered region (zero long-range order) by a line of
second-order phase transitions. Equation (\ref{split:hess})
defines the loci of the transition points in terms of a second
derivative of free energy of the system with respect of the order
parameters.
\end{thebibliography}
\end{document}